\documentclass[%
 reprint,
superscriptaddress,
 amsmath,amssymb,
 aps,
]{revtex4-1}

\usepackage{graphicx}
\usepackage{dcolumn}
\usepackage{bm}
\usepackage{color,soul}

\begin{document}


\title{Characteristic singular behaviors of nodal line materials emerging in orbital magnetic susceptibility and Hall conductivity}

\author{Ikuma Tateishi}  \email{i.tateishi@hosi.phys.s.u-tokyo.ac.jp}
 \affiliation{Department of Physics, The University of Tokyo, Bunkyo, Tokyo 133-0033, Japan}
\author{Viktor K\"{o}nye}
 \affiliation{Institute for Theoretical Solid State Physics, IFW Dresden, Helmholtzstrasse 20, 01069 Dresden, Germany}
\author{Hiroyasu Matsuura}
 \affiliation{Department of Physics, The University of Tokyo, Bunkyo, Tokyo 133-0033, Japan}
\author{Masao Ogata}
 \affiliation{Department of Physics, The University of Tokyo, Bunkyo, Tokyo 133-0033, Japan}



\date{\today}

\begin{abstract}
The bulk properties of nodal line materials have been an important research topic in recent years. In this paper, we study the orbital magnetic susceptibility and the Hall conductivity of nodal line materials using the formalism with thermal Green's functions and find characteristic singular behaviors of them. It is shown that, in the vicinity of the gapless nodal line, the orbital magnetic susceptibility shows a $\delta$-function singularity and the Hall conductivity shows a step function behavior in their chemical potential dependences. Furthermore, these singular behaviors are found to show strong field angle dependences corresponding to the orientation of the nodal line in the momentum space. These singular behaviors and strong angle dependences will give clear evidence for the presence of the nodal line and its orientation and can be used to experimentally detect nodal line materials.

\end{abstract}

\maketitle


\section{Introduction}

Topological semimetals in three-dimensional space have been extensively studied both theoretically and experimentally in the field of topological materials science.\cite{review2018a,review2018b}
They consist of mainly three kinds of phases: Weyl semimetals \cite{burkov2011weyl,halasz2012time,vafek2013dirac,huang2015weyl,jia2016weyl,yan2017topological,burkov2018weyl,armitage2018weyl}, Dirac semimetals \cite{kariyado2011three,young2012dirac,wang2012dirac,wang2013three,neupane2014observation,liu2014stable,borisenko2014experimental,yang2014classification}, and nodal line semimetals \cite{burkov2011topological,weng2015,kim2015dirac,yu2015topological,fang2015topological,yamakage2016,bian2016topological,fang2016topological,neupane2016observation,hu2016evidence,hirayama2017topological,kato2017a,kato2017b,takane2018observation,tateishi2018face,suzumura2018,suzumura2019,tateishi2020mapping}. Weyl semimetals have gapless points (Weyl points) in the momentum space and linear dispersion around the gapless points. These gapless points are monopoles of the Berry curvature and always appear in pairs with opposite chirality. Dirac semimetals also have gapless points (Dirac points) with linear dispersion, but the linear dispersive bands are doubly degenerated, like two overlapping Weyl points. In this case, the Dirac points must exist on high-symmetry lines in the momentum space and they are protected by some crystalline symmetry, such as rotational symmetry. The third kind of the topological semimetals or the nodal line semimetals that we study in this paper differ from Weyl or Dirac semimetals in that the gapless points are connected to a line (nodal line) in the three-dimensional momentum space. This nodal line is also protected by a crystalline symmetry or, if spin-orbit interactions are negligible, by time-reversal and space-inversion symmetries.

To confirm topological nature, angle-resolved photo-emission spectroscopy (ARPES) experiments have been a strong tool, which enables us to detect the surface states characteristic of the topological materials. For example, the Fermi arc \cite{wan2011topological,xu2015discovery,belopolski2016criteria,jia2016weyl}, which is one of the characteristic phenomena in Weyl semimetals, has been observed experimentally. 
The presence of the Fermi arc is topologically protected by the chirality of Weyl points. 
In contrast, there are no topologically protected surface states in the nodal line semimetals in the strict sense. Although the presence of drumhead surface states  \cite{burkov2011topological,bian2016drumhead,chan20163} has been reported, they are not topologically protected \cite{kargarian2016surface,fang2016topological,tateishi2020nodal}. The symmetries that guarantee the nodal lines are generally no longer present on the surface, and thus the drumhead surface states can depend on the surface configuration and the surface parameters, and even can be pushed out to the bulk spectrum by tuning the surface parameters. Therefore, the bulk properties for detecting the nodal lines, which do not rely on the surface states, are strongly demanded.

Quantum oscillations \cite{xiang2015angular,hu2016evidence,hu2016pi,zheng2016transport}, such as the Shubnikov-de Haas (SdH) oscillations, are a good experimental tool for this purpose. By observing the SdH oscillation and its phase offset, we can determine the dimensionality of the Fermi surface and the pocket type, usual parabolic dispersive pocket or singular linear dispersive pocket. The phase offsets are closely related to the formation of the Landau level in the high magnetic field region \cite{wang2016anomalous,yang2018quantum}. Although a detailed analysis is generally required to explicitly know the bulk dispersion \cite{wang2017quantum,he2014quantum,li2018rules}, the analysis of quantum oscillations and their field angular dependence is a powerful tool for investigating the features of topological semimetals, such as the structure of the Fermi surface and the structure of gapless points \cite{xiang2015angular,hu2016evidence}.

In the present paper, as alternative good bulk measurements, we study orbital magnetic susceptibility $\chi$ and Hall conductivity $\sigma_{xy}$, which enable us to confirm the existence of the nodal lines and to determine their directions in the momentum space. 
It is expected that $\chi$ and $\sigma_{xy}$ will have characteristic angle dependences: They will behave quite differently when the magnetic field is perpendicular to the plane formed by the nodal-line ring and when the magnetic field is parallel to the plane. 

There have been some theoretical studies on the magnetic susceptibility for the nodal line semimetals \cite{koshino2016magnetic,mikitik2016,mikitik2018,suzumura2019}.
However, the previous calculations assumed the local Weyl-type linear dispersion of two-dimensional momentum space at each point on the nodal line and obtained the total magnetic susceptibility approximately by integrating the local susceptibility along the nodal line. As a result, a $\delta$-function singularity has been observed when the magnetic field is parallel to the nodal line. 
In the present paper, we obtain $\chi$ exactly using the formalism with thermal Green's functions. Our result is consistent with the previous studies concerning the $\delta$-function singularity, but we find that there are additional contributions in $\chi$ that is, interestingly, very similar to the orbital magnetic susceptibility in three-dimensional massive Dirac electron systems.

The Hall conductivity in the weak magnetic field has been less understood compared with the magnetic susceptibility or the quantum oscillations. Only recently the quantum Hall effect due to the drumhead surface states has been discussed \cite{molina2018,zhaoCondmat}.
We will show that the Hall conductivity also shows a characteristic chemical potential dependence in the vicinity of the energy of the Dirac point, depending on the magnetic field direction. 
%

This paper is organized as follows:
In section II, we introduce a model Hamiltonian to describe nodal line materials.
In section III, we calculate the orbital magnetic susceptibility and its field angle dependence.
In section IV, we calculate the Hall conductivity in weak magnetic fields and its field angle dependence.
In section V, we give interpretations to the characteristic behavior of the obtained results by comparing them with the case of 2D Dirac electron systems.

\section{Hamiltonian of nodal line materials}

\begin{figure}
    \centering
    \includegraphics[width=8.5cm]{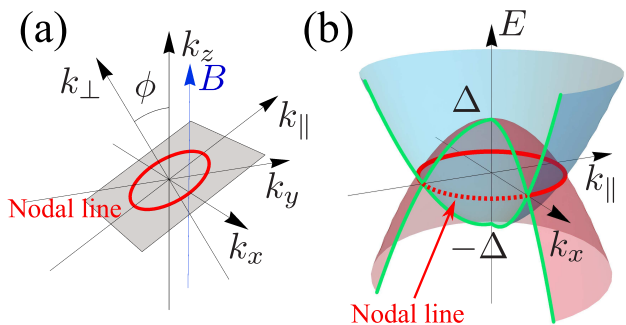}
    \caption{(a) Coordinate system used in the present paper and nodal line (red line) in the three-dimensional momentum space. The magnetic field $B$ is assumed to be parallel to the $k_z$-axis, and the nodal line is lying on the gray plane. (b) Band dispersion on the $k_x$-$k_\parallel$ plane when $k_\perp=0$. The nodal line (red line) is the intersection between the two parabolic bands (red and blue bands).}
    \label{fig:coord}
\end{figure}

In this section, we introduce a model Hamiltonian to describe nodal line materials. We assume that the spin-orbit coupling is negligible and construct a $\bm{k} \cdot \bm{p}$ perturbation Hamiltonian which hosts a ring-shape nodal line. The simplest Hamiltonian is given with two orbitals and the nodal line lies in a two-dimensional plane in the momentum space as shown in Fig.~\ref{fig:coord}.
We fix the magnetic field in the $z$-direction and assume that
the angle between the $k_z$-axis and the normal vector of the plane formed by the nodal line is $\phi$. 
Then, the Hamiltonian is given as
\begin{equation}
\label{eq:hamiltonian}
    H_{\bm{k}}= (a k_x^2 + b k_\parallel^2 - \Delta) \sigma_z + \nu k_\perp \sigma_x ,
\end{equation}
where $a, b, \nu$, and $\Delta$ are positive constants, $\sigma_{x}$ and $\sigma_z$ are Pauli matrices, and $k_\parallel$ and $k_\perp$ are defined as
\begin{equation}
    \left( 
    \begin{array}{c}
         k_\parallel \\
         k_\perp 
    \end{array} \right)
    =\left(
    \begin{array}{cc}
        \cos \phi & \sin \phi \\
        - \sin \phi & \cos \phi 
    \end{array} \right)
    \left(
    \begin{array}{c}
         k_y \\
         k_z 
    \end{array} \right).
\end{equation}

The eigenvalues of this Hamiltonian are
\begin{equation}
    E_{\pm} = \pm \sqrt{(a k_x^2 + b k_\parallel^2 - \Delta)^2 + \nu^2 k_\perp^2} = \pm \epsilon_{\bm k} .
    \label{eq:disp}
\end{equation}
Gapless points appear on the points where $E_+=E_-$ is satisfied, whose conditions are given by $k_\perp=0$ and $a k_x^2 + b k_\parallel^2 = \Delta$.  A ring-shape nodal line exists on the two-dimensional plane with $k_\perp=0$ as shown in Fig.~\ref{fig:coord}.

The present model (\ref{eq:hamiltonian}) is an extension of the previous model \cite{weng2015,kim2015dirac,yu2015topological,chan20163} with arbitrary angles relative to the magnetic field. 
It has been proposed for a low-energy effective Hamiltonian in several materials such as Cu$_3$ZnN \cite{kim2015dirac}, Ca$_3$P$_2$ \cite{chan20163}, TaTlSe$_2$ \cite{bian2016drumhead}, and CaAg$X$ ($X$=P, As) \cite{yamakage2016,takane2018observation}.

In the following sections, we calculate the orbital magnetic susceptibility $\chi$ and the Hall conductivity $\sigma_{xy}$ analytically using the thermal Green's functions. 
The thermal Green's function of the model (\ref{eq:hamiltonian}) is obtained as
\begin{eqnarray}
{\mathcal G}({\bm k}, i\epsilon_{n} ) &=&  \left[ i\epsilon_n - H_{\bm k} +\mu \right]^{-1} \nonumber \\
&=& \frac{1}{D} \left\{ (i\epsilon_n+\mu) \sigma_0 + A_{\bm k}\sigma_z + B_{\bm k}\sigma_x \right\},  \label{green1}
\end{eqnarray}
in a matrix form, where $\mu$ is the chemical potential, $\sigma_0$ is the $2 \times 2$ identity matrix, $A_{\bm k} =ak_x^2 + bk_{\parallel}^2 -\Delta$,  $B_{\bm k} =\nu k_{\perp}$, $D = (i \epsilon_n +\mu)^2- A_{\bm k}^2 - B_{\bm k}^2$, and 
$\epsilon_n$ is the Matsubara frequency, $\epsilon_n = (2n + 1)\pi k_{\rm B} T$ ($n \in \mathbb{Z}$). 
The energy eigenvalues are now written as $\pm \epsilon_{\bm k} = \pm \sqrt{A_{\bm k}^2+B_{\bm k}^2}$.

\section{Orbital Magnetic Susceptibility}

The research of orbital magnetic susceptibility has a long history since Landau and Peierls \cite{peierls,hebborn,fukuyamakubo,fukuyama1971theory,niu,piechon,ogatafukuyama,ogata,matsuura2016theory}. In particular, the problem of the large diamagnetism in Bi$_{1-x}$Sb$_x$ was resolved by Fukuyama and Kubo \cite{fukuyamakubo} by considering the interband effect of the magnetic field.
Then Fukuyama developed a general formula of the orbital susceptibility per volume \cite{fukuyama1971theory}
\begin{eqnarray}
\chi = \frac{e^2}{\hbar^2} \frac{k_{\rm B}T}{V} \sum_{n} \sum_{{\bf k}}{\rm Tr} \left[ {\mathcal G}\gamma_x{\mathcal G}\gamma_y{\mathcal G}\gamma_x{\mathcal G}\gamma_y \right],
\label{eq:Fukuyama}
\end{eqnarray}
where the spin degree of freedom has been included, $e$ is the electron charge ($e<0$), $V$ is the volume of the system, ${\mathcal G}:={\mathcal G}({\bm k},i\epsilon_{n})$ is an abbreviation of the thermal Green's function, 
and $\gamma_x$ and $\gamma_y$ are velocity operators in the $x$- and $y$-direction, respectively. 
The Fukuyama's formula (\ref{eq:Fukuyama}) is quite general and it has been applied to graphene \cite{fukuyama2007anomalous}, bismuth \cite{fuseya2014,fuseya2015transport}, and the Kane-Mele model \cite{ozakiogata}.
In particular, for graphene, the $\delta$-function singularity is reproduced, which was originally found by McClure \cite{mcclure}. This $\delta$-function singularity will be used later. 
It is to be noted that, in contrast to the previous studies \cite{koshino2016magnetic,mikitik2016}, it is not necessary to use the Landau levels, which are not always obtained analytically. 

When we apply the formula (\ref{eq:Fukuyama}) to the present model, the thermal Green's function is given in Eq.~(\ref{green1}) and the velocity operators are given by
\begin{eqnarray}
\gamma_x &=&\frac{\partial H_{\bm k}}{\partial k_x} = 2ak_x \sigma_x,  \label{vel3} \\
\gamma_y &=&\frac{\partial H_{\bm k}}{\partial k_y} = 2bk_{\parallel} \cos \phi \sigma_{z}
-\nu \sin \phi \sigma_x. \label{vel2}
\end{eqnarray}
Note that $\partial/\partial{k_y} = \cos \phi \partial/ \partial{k_\parallel} - \sin \phi \partial/\partial{k_\perp}$.
 
By substituting Eqs.~(\ref{green1}), (\ref{vel3}), and (\ref{vel2}) into Eq.~(\ref{eq:Fukuyama}), we find that the orbital magnetic susceptibility becomes
\begin{equation}
\chi = \chi_\perp \cos^2\phi  + \chi_\parallel \sin^2\phi, \label{eq:chi-phi}
\end{equation}
with
\begin{eqnarray}
\chi_{\perp} &=& \frac{e^2}{\hbar^2}k_{\rm B} T \sum_{n} \frac{\sqrt{ab}}{\nu}\int \frac{d{\bm p}}{(2\pi)^3}  \frac{32p_x^2p_{\parallel}^2}{D^2} \nonumber \\ 
&& \times \biggr( 1 + \frac{8A_{\bm p}^2}{D} + \frac{8A_{\bm p}^4}{D^2} \biggr), \label{Chi_intP} \\
\chi_{\parallel} &=& \frac{e^2}{\hbar^2}k_{\rm B} T \sum_{n} \sqrt{\frac{a}{b}}\nu\int  \frac{d{\bm p}}{(2\pi)^3}  \frac{8p_x^2}{D^2} \nonumber \\ 
&& \times \biggr(-1 + \frac{8A_{\bm p}^2 p_\perp^2}{D^2} \biggr),
\end{eqnarray}
where $p_x =\sqrt{a} k_x$, $p_{\parallel} =\sqrt{b} k_{\parallel}$, $p_{\perp} = \nu k_{\perp}$, and now $A_{\bm p}=p_x^2 +p_\parallel^2-\Delta$.
The term proportional to $\sin\phi \cos\phi$ has a $k_\parallel$-antisymmetric integrand and thus vanishes.

\begin{figure}
    \centering
    \includegraphics[width=8cm]{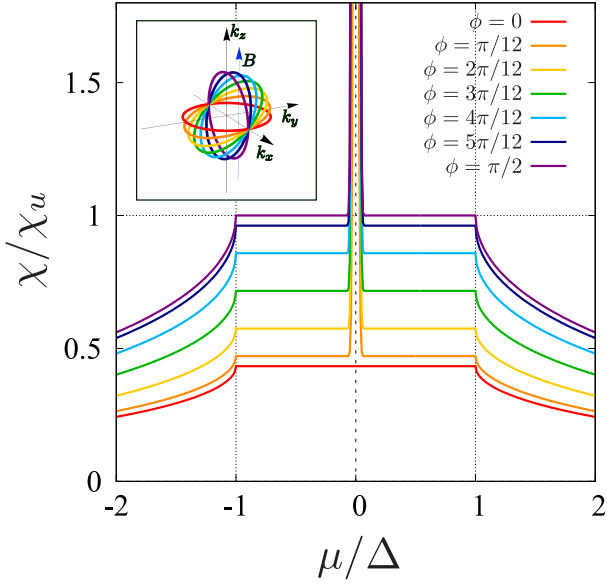}
    \caption{$\phi$ dependence of the orbital magnetic susceptibility. The nodal line orientations for each $\phi$ are shown in the inset.}
    \label{fig:chiphi}
\end{figure}

It is straightforward to perform the Matsubara summation and the $\bm p$-integral in $\chi_\perp$ using cylindrical coordinates, $p^2=p_x^2+p_\parallel^2$. At absolute zero ($T=0$), we obtain
\begin{eqnarray}
\chi_{\perp} =\left\{
 \begin{array}{ll}
-\frac{1}{6\pi^2}\frac{e^2}{\hbar^2}\frac{\sqrt{ab} \Delta}{\nu} \ln{\frac{2\Lambda}{\Delta}}, &\hspace{0.5cm}|\mu| \le \Delta, \\
-\frac{1}{6\pi^2}\frac{e^2}{\hbar^2}\frac{\sqrt{ab} \Delta}{\nu} \ln{\frac{2\Lambda}{|\mu| + \sqrt{\mu^2 - \Delta^2}}}, & \hspace{0.5cm} |\mu| \ge \Delta, \\
\end{array} \right.  \label{chi_z_anal}
\end{eqnarray} 
where $\Lambda$ is a cut-off energy. The details of the derivation are shown in Appendix A.
We find that the orbital susceptibility is constant for $ | \mu | \le \Delta$, while its value decreases as $| \mu |$ increases from $ | \mu | = \Delta$ as shown in Fig.~\ref{fig:chiphi}.
This chemical potential dependence is the same as that of three dimensional Dirac electron such as bismuth \cite{fukuyama2007anomalous,fuseya2015transport,fuseya2009interband}.

Similarly, the orbital susceptibility $\chi_\parallel$ for $T=0$ is calculated as follows.
\begin{eqnarray}
\chi_\parallel =\left\{
 \begin{array}{ll}
-\frac{1}{12\pi^2} \frac{e^2}{\hbar^2} \nu \sqrt{\frac{a}{b}} \ln{ \left( \frac{2\Lambda}{\Delta} \right) } +\chi^\prime, & \hspace{0.5cm}|\mu| \le \Delta, \\
-\frac{1}{12\pi^2} \frac{e^2}{\hbar^2} \nu \sqrt{\frac{a}{b}} \ln{ \frac{2\Lambda}{|\mu| + \sqrt{\mu^2 -\Delta^2}} }, & \hspace{0.5cm} |\mu| \ge \Delta, \\
\end{array} \right. \label{chi_x_anal}
\end{eqnarray} 
with
\begin{equation}
\chi^\prime
= - \frac{1}{3\pi} \frac{e^2}{\hbar^2} \nu \Delta \sqrt{\frac{a}{b}} \delta(\mu). \label{kaix2}
\end{equation}

Figure \ref{fig:chiphi} shows the obtained orbital magnetic susceptibility as a function of chemical potential $\mu$ for several choices of the angle $\phi$ ($\phi=\frac{\pi}{2}, \frac{5\pi}{12}, \cdots, \frac{\pi}{12}, 0$ from top to bottom). For convenience, $\chi$ is normalized with
\begin{equation}
    \chi_u = - \frac{1}{12\pi^2} \frac{e^2}{\hbar^2} \nu \sqrt{\frac{a}{b}} \ln \left( \frac{2\Lambda}{\Delta} \right),
\end{equation}
which is the constant value of $\chi_\parallel$ in $0 < |\mu|/\Delta < 1$. In the inset, the corresponding nodal line orientations are shown. In particular, according to Eq.~(\ref{eq:chi-phi}), the amplitude of the delta function $\chi^\prime$ at $\mu=0$ decreases as $\phi$ goes from $\pi/2$ ($\chi_\parallel$) to $0$ ($\chi_\perp$). 
This strong angle dependence of the magnetic susceptibility will give clear evidence for the presence of the nodal line and its orientation. 

At finite temperature, $\chi^\prime$ shows a characteristic temperature dependence 
\begin{equation}
\chi^\prime = - \frac{1}{12\pi} \frac{e^2}{\hbar^2} \nu \Delta \sqrt{\frac{a}{b}} \frac{1}{k_{\rm B}T} \frac{1}{\cosh^2 \frac{\mu}{2k_{\rm B}T}},
\end{equation}
instead of the $\delta$-function peak (see Eq.~(\ref{kaix2App})).

The singularity near $\mu=0$ is similar to that obtained in the two-dimensional massless Dirac electron systems \cite{mcclure,sharapov2004magnetic,fukuyama2007anomalous,koshinoando}, which will be discussed in detail in Section V.  
As shown in Eqs.~(\ref{chi_z_anal}) and (\ref{chi_x_anal}), there are additional contributions in $\chi$, which depend on the cut-off energy $\Lambda$. This behavior, in particular the cut-off energy dependence, is exactly the same as the orbital magnetic susceptibility in three-dimensional {\it massive} Dirac electron systems \cite{fukuyamakubo,fukuyama2007anomalous}. The origin of this behavior will be also discussed later.

\section{Hall Conductivity}

For studying the Hall conductivity, we use the microscopic formalism from Refs.~\cite{fukuyama1969theory,konye2020microscopic}, in which the conductivity is expressed using the retarded current-current correlation as
\begin{equation}
	\sigma_{xy}=\lim\limits_{\omega\to 0}\frac{ie^2}{\omega}\Pi^R_{xy}(\omega), 
       \label{eq:KuboHall}
\end{equation}
In the linear order of the magnetic field $B$, $\Pi^R_{xy}(\omega)$ is obtained by analytic continuation from \cite{fukuyama1969theory,konye2020microscopic}
\begin{align}
	\nonumber
	\Pi_{xy}(i\omega_\lambda)=-2i|e|B \frac{k_{\rm B}T}{\hbar^4 V}\sum\limits_{n,\bm{k}}&\mathrm{Tr}
	[\gamma_x {\mathcal G}^+ \gamma_y {\mathcal G} \gamma_x {\mathcal G} \gamma_y {\mathcal G} \\ 
	-&\gamma_x {\mathcal G}^+ \gamma_y {\mathcal G}^+ \gamma_x {\mathcal G}^+ \gamma_y {\mathcal G}] ,
	       \label{eq:KuboHall2}
\end{align}
where the spin degree of freedom has been included, ${\mathcal G}^+\equiv {\mathcal G}(\bm{k},i\epsilon_n+i\omega_\lambda)$ and $\omega_\lambda = 2\pi \lambda k_{\rm B} T$ with $\lambda$ being an integer is a Matsubara frequency representing the external frequency. 
The $i\omega_\lambda=\hbar\omega+i\eta$ substitution is made and the $\eta\to0$ limit is taken at the end.

In the eigenstate basis the Hall conductivity can be expressed as
\begin{align}
\sigma_{xy}&=-2B\frac{|e|^3}{\hbar^4V}\sum\limits_{\bm k}\sum\limits_{a,b,c,d} \gamma_{da}^x \gamma_{ab}^y \gamma_{bc}^x \gamma_{cd}^y C_{abcd} ,\\
C_{abcd}&=-\lim\limits_{\omega\to0}\frac{k_{\rm B}T}{\omega}\sum\limits_{n} {\mathcal G}_a^+ {\mathcal G}_d 
\left( {\mathcal G}_b {\mathcal G}_c-{\mathcal G}_b^+{\mathcal G}_c^+ \right) ,
\end{align}
where $\gamma_{da}^x$ represents the matrix element of $\gamma_x$ between the $d$-th and $a$-th band and the thermal Green's function of the $a$-th band is given by 
\begin{align}
    {\mathcal G}_a(\bm{k},i\epsilon_n)&=\frac{1}{i\epsilon_n-\varepsilon_a(\bm{k})+\mu+i\Gamma_a(\bm{k},i\epsilon_n)} .
\end{align}
For the transport properties, we need a finite scattering rate $\Gamma_a(\bm{k},i\epsilon_n)$, so that we use the eigenstate basis for $\sigma_{xy}$ in contrast to the case of $\chi$ in the previous section.
In the present model, we have only two bands and $\varepsilon_1({\bm k})=-\epsilon_{\bm k}$ and $\varepsilon_2({\bm k})=\epsilon_{\bm k}$. For the scattering rate we assume the simplest approximation where
\begin{equation}
    \Gamma_a(\bm{k},\varepsilon)=\mathrm{sign} \left( \mathrm{Im}(\varepsilon) \right) \Gamma ,
\end{equation}
where $\Gamma$ is constant. 

\subsection{Weak-scattering limit}

In the lowest order of the scattering rate ($\Gamma$) the Hall conductivity can be expressed as \cite{fukuyama1969theory,konye2020microscopic} (in the weak scattering limit this is the same as the Hall conductivity expressed using the Boltzmann transport theory)
\begin{equation}
    \begin{split}
        \sigma_{xy}^{\rm B} =& 2\frac{|e|^3 \tau^2 B}{\hbar^4 V} \sum_{\bm{k}}
%
\frac{\partial \epsilon_{\bm k}}{\partial{k_x}} \biggl\{
\frac{\partial \epsilon_{\bm k}}{\partial{k_x}} \frac{\partial^2 \epsilon_{\bm k}}{\partial{k_y}^2}
- \frac{\partial \epsilon_{\bm k}}{\partial{k_y}}
\frac{\partial^2 \epsilon_{\bm k}}{\partial{k_x} \partial{k_y}} \biggr\} \cr
&\times \left\{ f'(\epsilon_{\bm k})-f'(-\epsilon_{\bm k}) \right\},
        \label{eq:hc1}
    \end{split}
\end{equation}
where $f(\epsilon)$ is the Fermi distribution function defined by $f(\epsilon) =1/(e^{(\epsilon-\mu)/k_{\rm B}T} +1)$, and $\tau$ is the mean scattering time ($\Gamma=\hbar/2\tau$).
The subleading-order term with respect to the scattering rate is written in terms of the Berry curvature and orbital magnetic moment, but it vanishes in the present time-reversal symmetric case \cite{konye2020microscopic}.
Note that, as is well known in the case of graphene \cite{fukuyama2007anomalous,kobayashi2008hall}, this weak scattering limit is valid for $|\mu|\gtrsim \Gamma$ because we will have contributions in the order of $\Gamma/\mu$ in the small $\mu$-region. The effect of finite $\Gamma$ in the small $\mu$ region will be discussed in the next subsection. 

Using $\epsilon_{\bm k} = \sqrt{A_{\bm k}^2+B_{\bm k}^2}$ and 
$\partial/\partial{k_y} = \cos \phi \partial/ \partial{k_\parallel} - \sin \phi \partial/\partial{k_\perp}$, the Hall conductivity becomes
\begin{equation}
\sigma_{xy} = \sigma_{xy \perp}^{\rm B} \cos^2\phi + \sigma_{xy \parallel}^{\rm B} \sin^2\phi, 
\end{equation}
with
\begin{equation}
  \begin{split}
    \sigma_{xy \perp}^{\rm B} =& 2\frac{|e|^3 \tau^2 B}{\hbar^4} \frac{\sqrt{ab}}{\nu} \int \frac{d{\bm p}}{(2\pi)^3} \left\{ f'(\epsilon_{\bm p})-f'(-\epsilon_{\bm p}) \right\}
\frac{8p_x^2 A_{\bm p}^3}{\epsilon_{\bm p}^3}, \\
        \sigma_{xy \parallel}^{\rm B} =& 2\frac{|e|^3 \tau^2 B}{\hbar^4} \sqrt{\frac{a}{b}}\nu \int \frac{d{\bm p}}{(2\pi)^3} \left\{ f'(\epsilon_{\bm p})-f'(-\epsilon_{\bm p}) \right\}
\frac{4p_x^2 A_{\bm p}^2}{\epsilon_{\bm p}^3},
  \end{split}
  \label{eq:hc2}
\end{equation}
where $\epsilon_{\bm p}=\sqrt{A_{\bm p}^2+p_\perp^2}$.
As in the orbital magnetic susceptibility in the previous section, the term proportional to $\sin\phi \cos\phi$ has a $k_\parallel$-anti-symmetric integrand and thus vanishes.

At zero temperature, $f'(\epsilon_{\bm p})$ is explicitly written with the $\delta$-functions as 
\begin{equation}
\begin{split}
-\delta(\epsilon_{\bm p} -\mu) = -\frac{\mu \theta(\mu)}{\sqrt{\mu^2 - A_{\bm p}^2}} & \left[  \delta \left( p_\perp - \sqrt{\mu^2 - A_{\bm p}^2} \right) \right.\\
      &+ \left. \delta \left( p_\perp - \sqrt{\mu^2 - A_{\bm p}^2} \right) \right] ,
\end{split}
\end{equation}
where $\theta(\mu)$ is the Heviside function, i.e., $\theta(\mu)=1$ for $\mu>0$ and $0$ otherwise.
Using the cylindrical coordinates and $x=  p^2 -\Delta$ as in the case of orbital magnetic susceptibility, we obtain at $T=0$
\begin{widetext}
\begin{equation}\begin{split}
    \sigma_{xy \perp}^{\rm B} =& -2\frac{|e|^3 \tau^2 B}{\pi^2\hbar^4} \frac{\sqrt{ab}}{\nu} \frac{{\rm sign}(\mu)}{\mu^2}
    \int_{-\Delta}^{\infty} dx \frac{x^3(x+\Delta)\theta(\mu^2-x^2)}{\sqrt{\mu^2-x^2}} \\
&=\left\{
 \begin{array}{ll}
-\frac{3}{2} \frac{b}{\nu^2 \Delta} \sigma_u \mu^2 {\rm sign}(\mu), &\hspace{0.5cm}|\mu| \le \Delta, \\
-\frac{3}{2} \frac{b}{\nu^2 \Delta} \sigma_u \biggl[ \mu^2 \left\{ \frac{\pi}{2}+{\rm arctan}
\left( \frac{\Delta}{\sqrt{\mu^2-\Delta^2}}\right) \right\} + \frac{\Delta}{9} \left( 7+ \frac{2\Delta^2}{\mu^2}\right) \sqrt{\mu^2-\Delta^2} 
\biggr]{\rm sign}(\mu), & \hspace{0.5cm} |\mu| \ge \Delta, \\
\end{array} \right.  \\
\label{Hall_z_anal}
\end{split}\end{equation} 
and
\begin{equation}\begin{split}
    \sigma_{xy \parallel}^{\rm B} =& -\frac{|e|^3 \tau^2 B}{\pi^2\hbar^4} \nu \sqrt{\frac{a}{b}} \frac{{\rm sign}(\mu)}{\mu^2}
    \int_{-\Delta}^{\infty} dx \frac{x^2(x+\Delta)\theta(\mu^2-x^2)}{\sqrt{\mu^2-x^2}} \\
&=\left\{
 \begin{array}{ll}
-\sigma_u {\rm sign}(\mu), &\hspace{0.5cm}|\mu| \le \Delta, \\
-\sigma_u \biggl[ \left\{ \frac{\pi}{2}+{\rm arctan}
\left( \frac{\Delta}{\sqrt{\mu^2-\Delta^2}}\right) \right\}  + \frac{1}{3} \left( 4- \frac{\Delta^2}{\mu^2}\right) \frac{ \sqrt{\mu^2-\Delta^2} }{\Delta}
\biggr]{\rm sign}(\mu), & \hspace{0.5cm} |\mu| \ge \Delta, \\
\end{array} \right.
\label{Hall_x_anal}
\end{split}\end{equation}

\end{widetext}
where $\sigma_u$ is defined as 
\begin{equation}
    \sigma_u = \frac{|e|^3 \tau^2 B}{2\pi \hbar^4} \nu\Delta \sqrt{\frac{a}{b}}.
\end{equation}

\begin{figure}
    \centering
    \includegraphics[width=8cm]{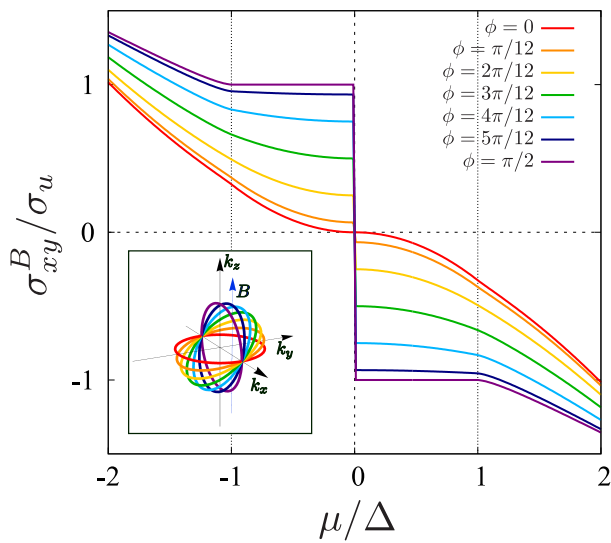}
    \caption{$\phi$ dependence of the Hall conductivity. The nodal line orientations for each $\phi$ are shown in the inset.}
    \label{fig:hallphi}
\end{figure}

Figure \ref{fig:hallphi} shows the obtained Hall conductivity as a function of chemical potential $\mu$ for some choices of the angle $\phi$ ($\phi=\frac{\pi}{2}, \frac{5\pi}{12}, \cdots, \frac{\pi}{12}, 0$ from left top to bottom, and from right bottom to top). In the inset, the corresponding nodal line orientations are shown. For a material with a fixed $\mu$, this strong angle dependence of $\sigma_{xy}$ will give clear evidence for the presence of the nodal line and its orientation. 

When the magnetic field is parallel to the nodal line ($\sigma_{xy \parallel}$, or $\phi=\pi/2$), the obtained Hall conductivity is constant at $0<|\mu|/\Delta<1$, but flips its sign at $\mu=0$ (violet line in Fig.~\ref{fig:hallphi}). This behavior is similar to the two-dimensional massless Dirac electron systems \cite{fukuyama2007anomalous}, which will be discussed in detail below. 
The step size at $\mu=0$ is $2\sigma_u$.

For example, if we choose the parameters as $\nu/\hbar \simeq 1.0 \times 10^6~\mathrm{[m/s]}$, $\Delta = 0.5 ~\mathrm{[eV]}$, $\tau=1.0\times10^{-13}~\mathrm{[s]}$, a transfer integral $t = 1.5 ~\mathrm{[eV]}$, and a lattice constant $L = 5.0 ~\mathrm{ \AA }$, 
then $b$ can be estimated as $b \sim \frac{L^2 t}{2} = 1.875 \times 10^{-19} ~\mathrm{[eV \cdot m^2]}$ and as a result, $\sigma_u$ becomes $\sigma_u \simeq 4.5\times10^{5}~\mathrm{[m^{-3} \cdot kg^{-1} \cdot s^3 \cdot A^2]}$, which is an experimentally observable value. In this assumption, the radius of the nodal line is roughly $0.26 \pi$. 

On the other hand, when the magnetic field is perpendicular to the nodal plane ($\sigma_{xy \perp}$, or $\phi=0$), the obtained Hall conductivity is approximately proportional to $-\mu^2 {\rm sign} (\mu)$. This behavior will be also discussed later.

\subsection{Effect of finite scattering near $\mu=0$}

\begin{figure}
    \centering
    \includegraphics[width=7cm]{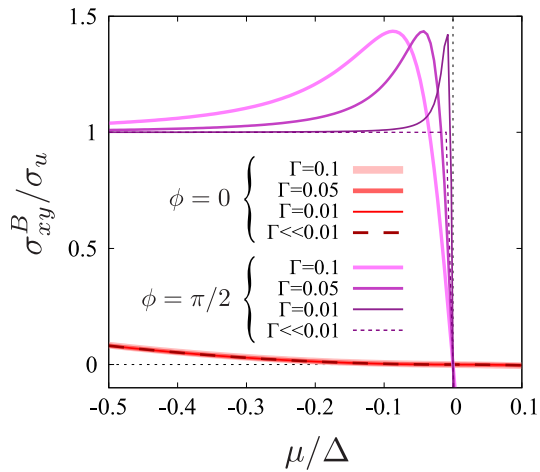}
    \caption{$\Gamma$ dependence of the Hall conductivity. Violet lines and red lines represent the cases of $\phi=\pi/2$ and $\phi=0$, respectively.}
    \label{fig:Gamma}
\end{figure}

As mentioned in the previous section, the weak-scattering limit is valid for $|\mu|\gtrsim \Gamma$. To obtain precisely the effects of the scattering rate in the small chemical potential region, we have to evaluate the Hall conductivity in Eqs.~(\ref{eq:KuboHall}) and (\ref{eq:KuboHall2}) at finite $\Gamma$ numerically. The obtained Hall conductivity can be expressed as (see Appendix. \ref{app:matsu})
\begin{align}
    \label{eq:sigxyGpa}
    \sigma_{xy\perp} &= \sigma_u \frac{b \Delta}{\nu^2} I_\perp (\tilde{\mu},\tilde{\Gamma}) ,\\
    \label{eq:sigxyGpe}
        \sigma_{xy \parallel} &= \sigma_u I_\parallel (\tilde{\mu},\tilde{\Gamma}) ,
\end{align}
where $\tilde{\Gamma}=\Gamma/\Delta$, $\tilde{\mu}=\mu/\Delta$, and $I_{\perp/\parallel}$ are dimensionless integrals in $x$ and $p_\perp$. Their explicit expressions are shown in Appendix \ref{app:matsu}. We evaluated these double integrals numerically and the results are shown in Fig.~\ref{fig:Gamma}.
%
%

As we can see at $|\mu| \gtrsim \Gamma$ we recover the analytic results of the previous section. At small chemical potentials the scattering rate does not really affect $\sigma_{xy \perp}$ in the perpendicular case. On the other hand, for the parallel case ($\sigma_{xy \parallel}$), we see a bump appearing in the plateau for small chemical potentials. The bump expands with increasing scattering rates. This result for the parallel case is very similar to the result obtained for graphene in Ref. \cite{fukuyama2007anomalous}, which will be discussed in the next section in detail. 

\section{Interpretation of the chemical potential dependences of $\chi$ and $\sigma_{xy}$}

\subsection{The parallel case ($\phi=\pi/2$)}

As shown in the previous sections, when the magnetic field is parallel to the plane where the nodal line exists, the chemical potential dependence of $\chi_\parallel$ has a $\delta$-function singularity and $\sigma_{xy \parallel}$ behaves like a step function. So let us consider this case first. 

\begin{figure}
    \centering
    \includegraphics[width=8cm]{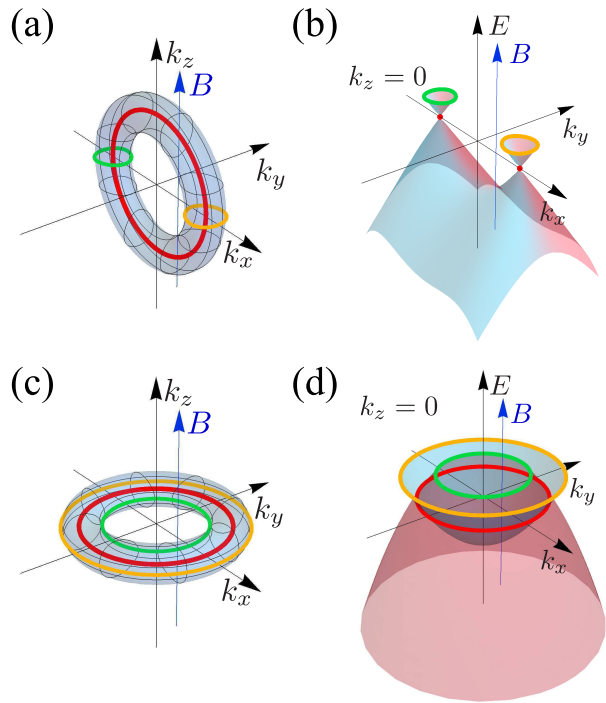}
    \caption{Fermi surfaces at $\mu=\Delta/2$ and band dispersions on the $k_z=0$
 (a) Fermi surface at $\mu=\Delta/2$ (blue surface) and the nodal line (red line) for the case of $\phi=\pi/2$. The green and yellow lines represent the Fermi surfaces on the $k_z=0$ plane. (b) Band dispersion on the $k_z=0$ plane. The red points represent the cross-sections of the nodal line, and the Fermi surfaces are shown by green and yellow lines at $E=\mu=\Delta/2$ on the $k_z=0$ plane.
 (c) Fermi surface at $\mu=\Delta/2$ (blue surface) and the nodal line (red line) for the case of $\phi=0$. The green and yellow lines represent the Fermi surfaces on the $k_z=0$ plane. (d) Band dispersion on the $k_z=0$ plane. The red line is the nodal line and the cross sections on $E=\mu$ are shown by green and yellow lines, which represent the Fermi surfaces on $k_z=0$. 
}
    \label{fig:fsband}
\end{figure}

In this case ($\phi=\pi/2$), the Fermi surfaces at $\mu=\Delta/2$ on the $k_z=0$ plane are shown in Fig.~\ref{fig:fsband}(a), which are the two separated rings. These rings are the cross-sections of two linear dispersive bands and they enclose the nodal line (Fig.~\ref{fig:fsband}(b)). 
Therefore, it is natural to interpret the chemical potential dependences of $\chi_\parallel$ and $\sigma_{xy \parallel}$ in terms of the two-dimensional massless Dirac electron systems, as was discussed in the previous studies on the magnetic susceptibility \cite{koshino2016magnetic,mikitik2016,mikitik2018,suzumura2019}.
In the present paper, we can compare the results obtained approximately by integrating the local susceptibility along the nodal line with the exact value obtained in this paper.

The two-dimensional massless Dirac electron system, or a model for graphene, is described by a Hamiltonian
\begin{equation}
    H = \gamma_x k_x \sigma_x + \gamma_y k_y \sigma_y.
    \label{eq:2DdiracH}
\end{equation}
In this model the orbital magnetic susceptibility has a $\delta$-function singularity \cite{mcclure,sharapov2004magnetic,fukuyama2007anomalous,koshinoando}
\begin{equation}
    \chi_{\rm Dirac}^{\rm 2D} = -\frac{e^2 \gamma_x \gamma_y}{3\pi\hbar^2} \delta(\mu),
    \label{eq:2DDiracChi}
\end{equation}
and the Hall conductivity behaves as \cite{fukuyama2007anomalous}. 
\begin{equation}
    \sigma_{\mathrm{Dirac}}^{\mathrm{2D}} = -\frac{|e|^3 \tau^2 B}{2 \pi \hbar^4} \gamma_x \gamma_y {\rm sign}( \mu ),
    \label{eq:2DDiracSigma}
\end{equation}
where the spin degrees of freedom has been taken into account. [Note that $\sigma_{\mathrm{Dirac}}^{\mathrm{2D}}$ can be understood from the classical form $\sigma_{xy, {\rm classical}}=-n_{\rm eff}|e|^3 \tau^2 B/m_{\rm eff}^2$ as follows. In the Dirac electron system, $n_{\rm eff}=2\pi k_{\rm F}^2$, while $m_{\rm eff}$ can be assumed to satisy $k_{\rm F}/m_{\rm eff}=\gamma_x$ or $\gamma_y$ (=constant), which means that $m_{\rm eff}$ is proportional to $k_{\rm F}$. Therefore, if we substitute $n_{\rm eff}=2\pi k_{\rm F}^2$ and $m_{\rm eff}^2 = k_{\rm F}^2/\gamma_x \gamma_y$ in $\sigma_{xy, {\rm classical}}$, we obtain Eq.~(\ref{eq:2DDiracSigma}).] In this section, we do not consider the bump appearing in the plateau for $\mu\sim 0$ (see Fig.~5), which will be understood similarly with this plateau value.  
%

Note that Eqs.~(\ref{eq:2DDiracChi}) and (\ref{eq:2DDiracSigma}) are for the two-dimensional systems and we need to transform them into contributions of the nodal line in the three-dimensional systems. Assume that there are $N_c$ independent layers of Dirac electron systems stacked three-dimensionally, each layer being separated by a distance $c$. Then the total magnetic susceptibility and the total Hall conductivity per volume become (using the length of the $c$-axis, $L_c=N_c c$)
\begin{eqnarray}
\chi_{\rm Dirac}^{\rm 3D} &=& \frac{N_c}{L_c} \chi_{\rm Dirac}^{\rm 2D}
=\frac{\chi_{\rm Dirac}^{\rm 2D}}{c}, \nonumber \\
\sigma_{\mathrm{Dirac}}^{\mathrm{3D}} &=& \frac{N_c}{L_c} \sigma_{\mathrm{Dirac}}^{\mathrm{2D}} = \frac{\sigma_{\mathrm{Dirac}}^{\mathrm{2D}}}{c}.
\end{eqnarray}
In this case, the length of the (straight) nodal line in the three-dimensional momentum space is $2\pi/c$. Therefore, the contributions of the nodal line per length should be
\begin{equation}
\chi_{\rm nodal/length}^{\rm 3D}=\frac{\chi_{\rm Dirac}^{\rm 2D}}{2\pi}, \qquad
\sigma_{\mathrm{nodal/length}}^{\mathrm{3D}} = \frac{\sigma_{\mathrm{Dirac}}^{\mathrm{2D}}}{2\pi}.
\end{equation}

\begin{figure}
    \centering
    \vspace{1cm}
    \includegraphics[width=6cm]{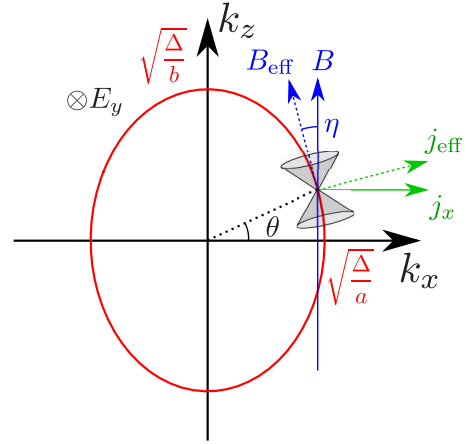}
    \caption{Nodal line (red line) of the $\phi=\pi/2$ case and locally defined 2D Dirac cone (gray cones).}
    \label{fig:2dDirac}
\end{figure}

In the present model, the nodal line forms an oval ring in the $k_x$-$k_z$ plane, and a point on the nodal line is expressed as (see Fig.~\ref{fig:2dDirac})
\begin{equation}
    \left(k_{0x}, k_{0y}, k_{0z} \right)=\left(\sqrt{\frac{\Delta}{a}} \cos \theta, 0, \sqrt{\frac{\Delta}{b}} \sin \theta \right).
\end{equation}
In the two-dimensional momentum space perpendicular to this nodal line, the band dispersion looks like a two-dimensional Dirac cone and thus the Hamiltonian is approximately written like Eq.~(\ref{eq:2DdiracH}) with properly chosen momenta. 
Actually, in the vicinity of the above point, by choosing $(k_{0x}+\delta k_x, k_{0y}+\delta k_y, k_{0z}+\delta k_z)$, the energy eigenvalues become
\begin{equation}
E_\pm = \pm \sqrt{(2ak_{0x} \delta k_x + 2bk_{0z} \delta k_z)^2+\nu^2 (\delta k_y)^2}.
\end{equation}
Therefore, we can see that the coefficients of $\bm k$ in Eq.~(\ref{eq:2DdiracH}), $\gamma_x$ and $\gamma_y$, are given as
\begin{equation}
\gamma_x = 2 \sqrt{\Delta}\sqrt{a \cos^2 \theta + b \sin^2 \theta}, \qquad
\gamma_y = \nu.
\end{equation}

The axis of the Dirac cone, which is normal to the two-dimensional momentum space, is 
\begin{equation}
{\bm t}=    \frac{1}{\sqrt{a \cos^2 \theta + b \sin^2 \theta}} \left(
  -\sqrt{b} \sin \theta, 0, \sqrt{a} \cos \theta \right),
\end{equation}
which is the tangent vector of the nodal line. Therefore, the angle $\eta$ between the magnetic field and the Dirac cone axis is
\begin{equation}
    \cos \eta = \frac{\sqrt{a} \cos \theta}{\sqrt{a \cos^2 \theta + b \sin^2 \theta}} .
\end{equation}

Now let us evaluate $\chi_\parallel$ by integrating the contribution of the nodal line, $\chi_{\rm nodal/length}^{\rm 3D}$, along the nodal line. Since the tangent vector $\bm t$ is not parallel to the magnetic field ($\parallel z$), the effective magnetic field is $B\cos \eta$. Furthermore, since the induced magnetic moment is also parallel to $\bm t$, we should integrate the $z$ component of this magnetic moment. The line integral along the nodal line using 
\begin{equation}
\sqrt{\frac{\Delta}{ab}} \sqrt{a \cos^2 \theta + b \sin^2 \theta}\  d\theta, 
\end{equation}
leads to
\begin{eqnarray}
M_z &=& \int_0^{2\pi}  \chi_{\rm nodal/length}^{\rm 3D} \cos^2 \eta B 
\sqrt{\frac{\Delta}{ab}} \sqrt{a \cos^2 \theta + b \sin^2 \theta} \ d\theta
 \nonumber \\
&=& -\int_0^{2\pi} \frac{e^2}{3\pi^2 \hbar^2} \delta(\mu)
 \nu\Delta\sqrt{\frac{a}{b}} B \cos^2\theta d\theta \nonumber\\
&=& -\frac{e^2}{3\pi \hbar^2} \delta(\mu) \nu\Delta\sqrt{\frac{a}{b}} B.
\label{eq:lineIntChi}
\end{eqnarray}
This exactly reproduces the obtained result $\chi^\prime$ in Eq.~(\ref{kaix2}).

We can see that the same argument holds for the Hall conductivity. As in the case of the magnetic susceptibility, the effective magnetic field is $B\cos\eta$. Furthermore, the induced Hall current $j_{\rm eff}$ is not parallel to the $x$-axis as shown in Fig.~\ref{fig:2dDirac}. Therefore, we need to integrate $j_x = j_{\mathrm{eff}} \cos \eta$ along the nodal line. As a result, we have the similar line integral as in Eq.~(\ref{eq:lineIntChi}):
\begin{eqnarray}
\langle j_x \rangle &=& \int_0^{2\pi}  \sigma_{\rm nodal/length}^{\rm 3D} \cos^2 \eta E_y 
\sqrt{\frac{\Delta}{ab}} \sqrt{a \cos^2 \theta + b \sin^2 \theta} \ d\theta
 \nonumber \\
&=& -\int_0^{2\pi} \frac{|e|^3 \tau^2 B}{2\pi^2 \hbar^4} {\rm sign}(\mu) \
 \nu\Delta\sqrt{\frac{a}{b}} E_y \cos^2\theta d\theta \nonumber\\
&=& -\frac{|e|^3 \tau^2 B}{2\pi \hbar^4} {\rm sign} (\mu) \ \nu\Delta\sqrt{\frac{a}{b}} E_y.
\label{eq:lineIntSigma}
\end{eqnarray}
This exactly reproduces the step of $\sigma_{xy \parallel}^{\rm B}$ at $\mu=0$, i.e., $\sigma_u$ obtained in the previous section.

The above arguments show that the $\delta$-function singularity in $\chi_\parallel$ and the plateau region in $\sigma_{xy \parallel}$ can be understood in terms of the nodal line. However, there is an additional contribution in $\chi_\parallel$ which depends on the energy cut-off. Since the two-dimensional massless Dirac electron system has only the $\delta$-functional singularity, this additional contribution can not be understood only from the nodal line. This will be due to the band dispersion that has not been taken into account in the two-dimensional massless Dirac model.

As for $\sigma_{xy \parallel}$, in the region of $|\mu|/\Delta>1$, the Hall conductivity is no longer constant and it decreases when $\mu>\Delta$ and increases when $\mu<-\Delta$. This is because the two rings of the Fermi surface touch each other at $\mu/\Delta=\pm1$ and they become a single large ring in $|\mu|/\Delta>1$. The single large ring encloses two Dirac points and thus the non-trivial property of the nodal line is not captured.

\subsection{The perpendicular case ($\phi=0$)}

The behavior of $\chi_\perp$ (Eq.~(\ref{chi_z_anal})) that is the same as the orbital magnetic susceptibility of bismuth can be understood from its Landau levels. In the present case, the energy eigenvalue under the magnetic field can be obtained analytically as $E=\pm E_{j, k_\perp}$ with
\begin{equation}
E_{j, k_\perp} = \sqrt{\left\{ \frac{2|e|\sqrt{ab} B}{\hbar} \left( j+\frac{1}{2} \right) -\Delta \right\}^2 + \nu^2 k_\perp^2},
\end{equation}
where $j=0, 1, \cdots$ is the Landau level index. The grand potential is expressed as
\begin{equation}
\Omega = -2 k_{\rm B}T \frac{|e| BL^2}{2\pi \hbar} \sum_{\pm, k_\perp} \sum_{j=0}^\infty \ln \left\{ 1+e^{-\beta(\pm E_{j, k_\perp}-\mu)} \right\}, 
\end{equation}
where the prefactor $|e|BL^2/2\pi \hbar$ represents the degeneracy of each Landau level. 
In the small magnetic field region, the summation over the Landau level $j$ can be estimated by using the Euler-MacLaurin expansion for a smooth function $F(\lambda)$ and for a large $N$:
\begin{equation}
\begin{split}
& \frac{1}{N} \sum_{j=n_1}^{n_2} F\left( \frac{j}{N} \right) = \int_{\frac{n_1-1/2}{N}}^{\frac{n_2+1/2}{N}} F(\lambda) d\lambda \\
&+ \frac{1}{24N^2} \left\{ F' \left( \frac{n_1-\frac{1}{2}}{N} \right) - F' \left( \frac{n_2+\frac{1}{2}}{N} \right) \right\} + O \left( \frac{1}{N^3} \right),
\label{eq:Euler}
\end{split}
\end{equation}
where we can assume $N=1/B$ and $x=Bj$. Then, after some algebra, 
we obtain the grand potential $\Omega$ as
\begin{equation}
\begin{split}
\Omega =& -2 k_{\rm B}T \frac{|e| L^2}{2\pi \hbar} \sum_{\pm, k_\perp} \biggl[ 
\int_0^\infty \ln \left\{ 1+e^{-\beta ( \pm E(x,k_\perp)-\mu ) } \right\} dx \\
& \pm \frac{B^2}{24} \frac{2|e|\Delta \sqrt{ab}}{k_{\rm B}T\hbar \sqrt{\Delta^2+ \nu^2 k_\perp^2}} 
f \left( \pm \sqrt{\Delta^2 + \nu^2 k_\perp^2} \right) \biggr],
\label{eq:Omega}
\end{split}
\end{equation}
with $E(x, k_\perp) = \sqrt{ (2|e|\sqrt{ab}x/\hbar -\Delta )^2 + \nu^2 k_\perp^2}$.
The first term in Eq.~(\ref{eq:Omega}) represents the grand potential at $B=0$. From the second term, we obtain
\begin{equation}
\begin{split}
\chi &= -\frac{\partial^2 \Omega}{\partial B^2} \\
&= \frac{e^2 L^2}{6\pi \hbar^2} \sum_{\pm, k_\perp} \left[ 
\pm \frac{\Delta \sqrt{ab}}{\sqrt{\Delta^2+ \nu^2 k_\perp^2}} 
f \left( \pm \sqrt{\Delta^2 + \nu^2 k_\perp^2} \right) \right],
\end{split}
\end{equation}
When we perform the $k_\perp$ integral at $T=0$, we reproduce the result in Eq.~(\ref{chi_z_anal}).

For bismuth, we have Landau levels as\cite{,fuseya2015transport}
\begin{equation}
E_{j, k_\perp}^{\rm Bi} = \sqrt{\Delta^2 + \frac{2|e|\gamma_x \gamma_y B}{\hbar} \left( j+\frac{1}{2} + \frac{\sigma_z}{2} \right) + \gamma_z^2 k_\perp^2},
\end{equation}
where $\sigma_z$ takes values $\pm 1$. Although there are some differences between the present case and bismuth, the Euler-MacLaurin expansion gives a similar grand potential in both cases, which leads to our results that $\chi_\perp$ in the present model has the same $\mu$-dependence as the orbital magnetic susceptibility in bismuth. The main reason for this coincidence is that the magnetic susceptibility is determined by the term $F'((n_1-1/2)/N)$ in Eq.~(\ref{eq:Euler}) that is related to the first Landau level with $j=0$, and that the energy of the first Landau level is $\sqrt{\Delta^2 + \gamma_x^2 k_z^2}+O(B)$ in both cases of the present case and bismuth. 

Next we discuss $\sigma_{xy\perp}$ in Fig.~\ref{fig:hallphi}.
Its $\mu$ dependence is simply explained by the structure of the Fermi surface. In the weak scattering limit, i.e., in the semi-classical picture, the Hall effect is discussed within a two-dimensional momentum space perpendicular to the magnetic field. At the same time, at zero temperature only the contributions from the Fermi surface are to be taken into account. Therefore, the structure of the intersection of the Fermi surface and $k_z=\mathrm{const.}$ plane determines the behavior of the Hall conductivity.
For $\phi=0$, the Fermi surfaces on a $k_z=\mathrm{const.}$ plane are two concentric rings (Fig.~\ref{fig:fsband}(c)). These concentric rings are the cross-sections of two parabolic bands and they do not enclose the nodal line (Fig.~\ref{fig:fsband}(d)). Therefore, this Fermi surface structure gives the free-electron-like Hall conductivity as shown in the red line in Fig.~\ref{fig:hallphi}.

To make more quantitative interpretation, let us use again the classical Hall conductivity $\sigma_{xy, {\rm classical}}=-n_{\rm eff}|e|^3 \tau^2 B/(m_{x, {\rm eff}} m_{y, {\rm eff}})$. In the present case, we can assume $m_{x, {\rm eff}}=\hbar^2 /2a$ and $m_{y, {\rm eff}}=\hbar^2 /2b$ and that $n_{\rm eff}$ is estimated from the volume of the Fermi surface. Let us consider the case with $0<\mu<\Delta$. In this case the yellow line in Fig.~\ref{fig:fsband}(d) is the electron Fermi surface and the green line is the hole Fermi surface. Taking into account the $k_\perp$-direction, $n_{\rm eff}$ that is electron density minus hole density becomes
\begin{equation}
\begin{split}
n_{\rm eff} =& \frac{2}{(2\pi)^3} \int_{-\frac{\mu}{\nu}}^{\frac{\mu}{\nu}} dk_\perp \int dk_x dk_\parallel \\
&\times \left[ \theta \left( \sqrt{\mu^2-\nu^2 k_\perp^2}-A_{\bm k} \right) -\theta \left( A_{\bm k}-\sqrt{\mu^2-\nu^2 k_\perp^2} \right) \right] \\
=&\frac{1}{4\pi^2} \int_{-\frac{\mu}{\nu}}^{\frac{\mu}{\nu}}dk_\perp \\
& \times \left[ \frac{\Delta+\sqrt{\mu^2-\nu^2k_\perp^2}}{\sqrt{ab}} - \frac{\Delta-\sqrt{\mu^2-\nu^2k_\perp^2}}{\sqrt{ab}} \right] \\
=&\frac{\mu^2}{4\pi\nu \sqrt{ab}},
\end{split}
\end{equation}
where $\theta(x)$ is the Heviside step function and $A_{\bm k}=ak_x^2 +bk_\parallel^2-\Delta$. The first term corresponds to the electron density and the second term to the hole density. Substituting of $n_{\rm eff}$ into $\sigma_{xy, {\rm classical}}$, we obtain
\begin{equation}
\sigma_{xy, {\rm classical}}=-\frac{|e|^3\tau^2 B}{\pi \hbar^4} \frac{\sqrt{ab}}{\nu} \mu^2 ~\left(= 2 \frac{b}{\nu^2 \Delta} \sigma_u \mu^2 \right),
\end{equation}
which reproduces the $\mu^2$-dependence of the exactly obtained result $\sigma_{xy \perp}$ in Eq.~(\ref{Hall_z_anal}). The difference exists only in the numerical prefactor, $3/2 \rightarrow 2$.

Similarly, for the case with $\mu>\Delta$, we obtain
\begin{equation}
\begin{split}
n_{\rm eff}
=& \frac{1}{4\pi^2} \int_{-\frac{\mu}{\nu}}^{\frac{\mu}{\nu}}dk_\perp \biggl[ 
\frac{\Delta+\sqrt{\mu^2-\nu^2k_\perp^2}}{\sqrt{ab}} \\
&-\frac{\Delta-\sqrt{\mu^2-\nu^2k_\perp^2}}{\sqrt{ab}} 
\theta \left( \Delta-\sqrt{\mu^2-\nu^2k_\perp^2}\right) \biggr] \\
=& \frac{1}{4\pi^2\nu \sqrt{ab}} \biggl\{ \mu^2 \left( \frac{\pi}{2}+\tan^{-1}\frac{\Delta}{\sqrt{\mu^2-\Delta^2}}  \right) \\
&+ \Delta\sqrt{\mu^2-\Delta^2} \biggr\}.
\end{split}
\end{equation}
From this $n_{\rm eff}$,  $\sigma_{xy \perp}$ in Eq.~(\ref{Hall_z_anal}) is reasonably reproduced.

\section{Summary}
We have calculated the chemical potential dependence and magnetic field angle dependence of the orbital magnetic susceptibility and the Hall conductivity in nodal line materials. These quantities show characteristic singular behaviors in the chemical potential dependence, which is attributed to the non-trivial gapless structure of the bulk band dispersion. They also show a strong field angle dependence corresponding to the orientations of the nodal line. These results allow us to detect the presence of nodal lines and to determine their orientations in the momentum space using bulk properties that are independent of the surface state.

\section*{Acknowledgement}
This work was supported by Grants-in-Aid for Scientific Research from the Japan Society for the Promotion of Science (Grant No. JP18H01162), and by JST-Mirai Program (Grant No. JPMJMI19A1).
I.T. was supported by the Japan Society for the Promotion of Science through the Program for Leading Graduate Schools (MERIT).

\vspace{5cm}
\appendix

\section{Matsubara summation and momentum integrals in orbital magnetic susceptibility} 

Using cylindrical coordinates, the susceptibility $\chi_{\perp}$ in Eq.~(\ref{Chi_intP}) is expressed as
\begin{widetext}
\begin{equation}
\begin{split}
\chi_{\perp} &= \frac{e^2}{\hbar^2}k_{\rm B} T \sum_{n}\frac{1}{2\pi^2}\frac{\sqrt{ab}}{\nu}\int_{-\infty}^{\infty} dp_{\perp} \int_{-\Delta}^{\infty}dx \frac{(x + \Delta)^2}{(x^2 + y^2)^2} \biggr\{ 1 - \frac{8x^2}{x^2 +y^2} + \frac{8x^4}{(x^2 +y^2)^2} \biggr\} \\
 &= -\frac{e^2}{\hbar^2}k_{\rm B} T \sum_{n}\frac{1}{2\pi^2}\frac{\sqrt{ab}}{\nu}\int_{-\infty}^{\infty} dp_{\perp}  \frac{\Delta}{3(\Delta^2 +y^2)} \\
 &= -\frac{1}{2\pi^2}\frac{e^2}{\hbar^2}\frac{\sqrt{ab}}{\nu}\oint\frac{dz}{2\pi i}f(z) \int_{-\infty}^{\infty} dp_{\perp} \frac{\Delta}{3(z^2 - p_{\perp}^2 - \Delta^2 )} \\
 &= \frac{1}{2\pi^2}\frac{e^2}{\hbar^2}\frac{\sqrt{ab}}{\nu} \int_{-\infty}^{\infty} dp_{\perp} \frac{\Delta}{6\epsilon_z}\left\{ f(\epsilon_z)- f(-\epsilon_z) \right\}, 
\end{split}
\end{equation}
where $x=  p^2 -\Delta$, $y^2 = -(i\epsilon_n+\mu)^2 + p_{\perp}^2$, $\epsilon_z =\sqrt{\Delta^2 + p_{\perp}^2}$, and $f(z)$ is Fermi distribution function defined by $f(z) =1/(e^{(z-\mu)/k_{\rm B}T} +1)$, respectively.
At $T=0$, the $p_\perp$-integral leads to Eq.~(\ref{chi_z_anal}).

Similarly, the susceptibility of $\chi_\parallel$ is calculated as
\begin{equation}
\begin{split}
\chi_\parallel &= \frac{e^2}{\hbar^2}k_B T\sum_{n} \frac{\nu}{2\pi^2}\sqrt{\frac{a}{b}}\int_{-\infty}^{\infty} dp_{\perp} \int_{-\Delta}^{\infty} dx \frac{x + \Delta}{(x^2 + y^2)^2} \left\{ -1 + \frac{8x^2 p_{\perp}^2}{(x^2 + y^2)^2} \right\} \\
         &= \frac{e^2}{\hbar^2}k_B T\sum_{n} \frac{\nu}{2\pi^2}\sqrt{\frac{a}{b}}\int_{-\infty}^{\infty} dp_{\perp} \left\{ -\frac{y^2-p_\perp^2}{2y^4} + \frac{p_{\perp}^2(y^2 - \Delta^2)}{6y^2 (\Delta^2 + y^2)^2} 
         +\frac{\Delta(y^2 -p_{\perp}^2)}{2y^4} \int_{\Delta}^{\infty} \frac{dx}{x^2 + y^2}  - \frac{\pi \Delta(y^2 -p_{\perp}^2)}{2y^5} \right\}. \label{kaix1}
\end{split}
\end{equation}
The last term of Eq.~(\ref{kaix1}) becomes
\begin{eqnarray}
\chi^\prime &\equiv& -\frac{e^2}{\hbar^2}k_B T\sum_{n} \frac{\nu}{2\pi^2}\sqrt{\frac{a}{b}}\int_{-\infty}^{\infty} dp_{\perp} \frac{\pi \Delta(y^2 -p_{\perp}^2)}{2y^5} \nonumber \\
&=& -\frac{e^2}{\hbar^2}k_B T\sum_{n} \frac{\nu}{2\pi^2}\sqrt{\frac{a}{b}} \frac{2\pi}{3}\frac{\Delta}{\epsilon_n^2}  \nonumber  \\
&=& -\frac{1}{3\pi} \frac{e^2}{\hbar^2} \nu \sqrt{\frac{a}{b}} \oint \frac{dz}{2\pi i} f(z) \frac{\Delta}{z^2} \nonumber  \\
&=& \frac{\nu \Delta}{3\pi}\frac{e^2}{\hbar^2}\sqrt{\frac{b}{a}} f^{\prime}(0). \label{kaix2App}
\end{eqnarray}
At $T = 0$, $\chi^\prime$ gives a $\delta$-function singularity because $f^\prime(0) = -\delta(\mu)$.

The remaining terms of Eq.~(\ref{kaix1}) can be integrated as
\begin{equation}
\begin{split}
\chi_\parallel - \chi^\prime &= \frac{1}{2\pi^2} \frac{e^2}{\hbar^2} \nu \sqrt{\frac{a}{b}} \int_{-\infty}^{\infty}dp_{\perp} \sum_{\pm} \left[ \frac{p_\perp^2 f^\prime(\pm \epsilon_z)}{12\epsilon_z^2} \mp \frac{p_\perp^2 (\epsilon_z^2+\Delta^2)  f(\pm \epsilon_z)}{12\epsilon_z^3 \Delta^2} \pm \Delta \int_{\Delta}^{\infty} dx \frac{\sqrt{x^2 + p_{\perp}^2} f \left( \pm \sqrt{x^2 + p_{\perp}^2} \right) }{4x^4} \right] \\
&= \frac{1}{24\pi^2} \frac{e^2}{\hbar^2} \nu \sqrt{\frac{a}{b}} \int_{-\infty}^{\infty}dp_{\perp}  \left[ \frac{\epsilon_z}{\Delta^2} \left\{ -f(\epsilon_z) + f(-\epsilon_z)\right\} + 3\Delta \int_{\Delta}^{\infty} dx \frac{\sqrt{x^2 +p_{\perp}^2}}{x^4} \left\{  f \left( \sqrt{x^2 + p_{\perp}^2} \right) -  f \left( -\sqrt{x^2 + p_{\perp}^2} \right) \right\} \right].
\end{split}
\end{equation}
At $T=0$, the $p_\perp$- and $x$-integral lead to Eq.~(\ref{chi_x_anal}).

\section{Matsubara summation for the finite scattering rate case}

\label{app:matsu}
We calculate the Hall conductivity in Eqs.~(\ref{eq:KuboHall}) and (\ref{eq:KuboHall2}) at finite $\Gamma$ numerically. The Matsubara summation is evaluated by transforming the summation to an integral using the Fermi distribution function \cite{Bruus2004,konye2020microscopic}. 
\begin{equation}\begin{split}
\sigma_{xy}^\perp &= 2\frac{|e|^3 B}{\hbar^2 V} \sum_{\bm k} 32 a^2 b k_x^2 {\rm Re} \left[ 
\int\frac{d\varepsilon}{2\pi i} f'(\varepsilon) A_{\bm k} \left( \frac{1}{2D_R^2} + \frac{2A_{\bm k}^2}{3D_R^3} 
- \frac{\varepsilon^2 +\Gamma^2 +A_{\bm k}^2 - B_{\bm k}^2}{D_R^2 D_A} \right) \right], \\
\sigma_{xy}^\parallel &= 2\frac{|e|^3 B}{\hbar^2 V} \sum_{\bm k} 16 a^2 k_x^2 \nu^2 {\rm Re} \left[ 
\int\frac{d\varepsilon}{2\pi i} f'(\varepsilon) \left( \frac{1}{2D_R^2} + \frac{2A_{\bm k}^2}{3D_R^3} 
- \frac{\varepsilon^2 +\Gamma^2 +A_{\bm k}^2 - B_{\bm k}^2}{D_R^2 D_A} \right)\right],
\end{split}\end{equation}
where $D_R = (\varepsilon+i\Gamma)^2- A_{\bm k}^2 - B_{\bm k}^2$ and $D_A = (\varepsilon-i\Gamma)^2- A_{\bm k}^2 - B_{\bm k}^2$

At zero temperature the integration can be done analytically using $f'(\varepsilon)=-\delta(\varepsilon-\mu)$. Then we use cylindrical coordinates and integrate the azimuth variable analytically. For the remaining two integrals we make the expressions dimensionless and get
\begin{equation}
    \sigma_{xy \perp} = \sigma_u \frac{b \Delta}{\nu^2} I_\perp(\tilde{\mu},\tilde{\Gamma}),
\qquad    \sigma_{xy \parallel} = \sigma_u I_\parallel(\tilde{\mu},\tilde{\Gamma}) ,
\end{equation}
where $\tilde{\Gamma}=\Gamma/\Delta$, $\tilde{\mu}=\mu/\Delta$, and
\begin{align}
    I_\perp &= - \frac{64}{3\pi^2}\tilde{\mu}\tilde{\Gamma}^5\int\limits_{-1}^\infty dK \int\limits_{-\infty}^{\infty} dk ~\frac{g_\parallel(\tilde{\mu},\tilde{\Gamma},K,k)}{h(\tilde{\mu},\tilde{\Gamma},K,k)} ,\\
    I_\parallel &= - \frac{32}{3\pi^2}\tilde{\mu}\tilde{\Gamma}^5\int\limits_{-1}^\infty dK \int\limits_{-\infty}^{\infty} dk ~\frac{g_\perp(\tilde{\mu},\tilde{\Gamma},K,k)}{h(\tilde{\mu},\tilde{\Gamma},K,k)} ,\\
    \nonumber
    g_\perp&=K(K+1)\bigg[3\left(k^2+K^2+\tilde\Gamma^2\right)^2 +2 \left(-3k^2+5 K^2+3\tilde\Gamma^2\right) \tilde\mu^2+3\tilde\mu^4\bigg] ,\\
    \nonumber
    g_\parallel&=g_\perp/K\\
    h&=\left[\left(k^2+K^2+\tilde{\Gamma}^2\right)^2-2 \left(k^2+K^2-\tilde{\Gamma}^2\right) \tilde{\mu}^2+\tilde{\mu}^4\right]^3,
\end{align}
with $K=x/\Delta$ and $k=p_\perp/\Delta$. 


\end{widetext}

Note that the same expression can be achieved starting from Eq. (\ref{eq:KuboHall2}). The matrix elements of the velocity operators can be expressed as
\begin{equation}
    \mathrm{Tr}\left[\gamma_x Q_a \gamma_y Q_b \gamma_x Q_c \gamma_y Q_d\right],
\end{equation}
where $Q_a$ are the projection operators of the Hamiltonian. The projection operators can be calculated using the Frobenius covariant \cite{horn1994topics,Cserti2010}:
\begin{equation}
    Q_a = \prod\limits_{\substack{b\\a\neq b}} \frac{H-E_b}{E_a-E_b}.
\end{equation}
In the present model there are only two bands with $\pm \epsilon_{\bm k}$. 

\bibliography{reference}

\end{document}